\documentclass[prl,aps,showpacs,twocolumn]{revtex4}
\usepackage{amsfonts}
\usepackage{amsmath}
\usepackage{amssymb}
\usepackage[dvips]{graphicx}

\begin{document}

\title{Direct measurement of concurrence for atomic two-qubit pure states}
\date{\today}
\author{G. Romero$^1$, C.E. L\'{o}pez$^1$, F. Lastra$^1$, E. Solano$^{2,3}$
, and J.C. Retamal$^1$}
\affiliation{$^1$Departamento de F\'{\i}sica, Universidad de Santiago de Chile, Casilla
307 Correo 2, Santiago, Chile \\
$^2$Physics Department, ASC, and CeNS, Ludwig-Maximilians-Universit\"at,
Theresienstrasse 37, 80333 Munich, Germany \\
$^3$Secci\'on F\'{\i}sica, Departamento de Ciencias, Pontificia
Universidad Cat\'olica del Per\'u, Apartado 1761, Lima, Peru}

\begin{abstract}
We propose a general scheme to measure the concurrence of an
arbitrary two-qubit pure state in atomic systems. The protocol is
based on one- and two-qubit operations acting on two available
copies of the bipartite system, and followed by a global qubit
readout. We show that it is possible to encode the concurrence in
the probability of finding all atomic qubits in the ground state.
Two possible scenarios are considered: atoms crossing 3D microwave
cavities and trapped ion systems.
\end{abstract}

\maketitle

Quantum entanglement is a key resource for quantum information and
quantum computation~\cite{nielsen00}. This intriguing property lies
at the heart of the Einstein-Podolsky-Rosen paradox~\cite{EPR}.
Entangled states have been implemented in different physical setups,
for example, in photons~\cite{tittel01}, massive particles like
trapped ions~\cite{leibfried03}, nuclear magnetic
resonance~\cite{vandersypen04}, atoms in cavities~\cite{raimond01},
quantum dots~\cite{hanson06}, among others. On the other hand, the
quantification of the degree of entanglement for an arbitrary number
of qubits is still an open problem in quantum
information~\cite{mintert05}. Arguably, the most valuable
entanglement measure is the entanglement of formation
(EOF)~\cite{hill97}, which quantifies the minimal cost needed to
prepare a certain quantum state in terms of EPR pairs. Many efforts
have been devoted to the derivation of the EOF through analytical
and numerical approaches. In an important contribution it has been
shown that EOF $E_{f}(\rho)$ for an arbitrary two-qubit state $\rho$
can be defined in terms of an exactly calculable quantity: the
concurrence $C$~\cite{wootters98}. This quantity can be defined as
$C(\rho )=\max \{0,\lambda _{1}-\lambda _{2}-\lambda _{3}-\lambda
_{4}\}$, where the $\lambda _{i}$'s are square roots in decreasing
order of the eigenvalues of matrix $\rho \tilde{\rho}$ with
$\tilde{\rho}=\sigma _{y}\otimes \sigma _{y}\rho ^{\ast }\sigma
_{y}\otimes \sigma _{y}$, $\sigma_y$ being the usual Pauli operator.
Remarkably, for a pure state this concurrence is reduced to the
simple expression

\begin{equation}
\label{pureconcurrence} C(|\psi _{i}\rangle )=|\langle \psi |\sigma
_{y}\otimes \sigma _{y}|\psi ^{\ast }\rangle | .
\end{equation}

A straightforward method for measuring entanglement would be a
complete tomographic reconstruction of the quantum state
\cite{White99}. In this case, the reconstruction of a two-qubit
state requires the readout of 15 parameters. Additionally,
theoretical proposals based on entanglement Witness operator
\cite{Ekert1}, positive maps \cite{horodecki1}, and two-particle
interference~\cite{Zhou}, have been introduced. Recently, the direct
measurement of concurrence of a two-photon entangled state was
implemented in the lab~\cite{walborn}. This experiment is based on
the fact that the concurrence information of a two-qubit pure state
is encoded in the probability of observing the two copies of the
first subsystem in an antisymmetric state~\cite{mintert}. Without
any doubt, it would be desirable to translate these ideas to the
case of matter qubits where diverse physical setups have reached
high level of quantum control.

In this work, we propose a method to measure the concurrence of a
two-qubit pure state in matter qubits. The proposed technique relies
on the availability of two copies of the bipartite state and the
direct measurement of the occupation probability of the collective
state of both copies. We illustrate this protocol with two examples,
Rydberg atoms crossing 3D microwave cavities~\cite{raimond01} and
confined ions in a linear Paul trap~\cite{leibfried03}.

The central idea of this proposal is the transformation of the
separable state of two copies into a state where the value of the
concurrence will be loaded in the probability amplitude to have all
the qubits in the ground state. The required operations are
$\sigma_y$ unitaries and local rotations $R$, as well as a
controlled-not gate (CNOT), followed by a global measurement of all
four qubits. In Fig.~1 we present a quantum circuit describing the
proposed protocol. Here, the first two channels stand for the
entangled state we want to measure, the third and fourth channel
denote the copy of the two-qubit state. Finally, the measurement is
produced through the detection of all qubits in the ground state.

\begin{figure}[b]
\includegraphics[width=55mm]{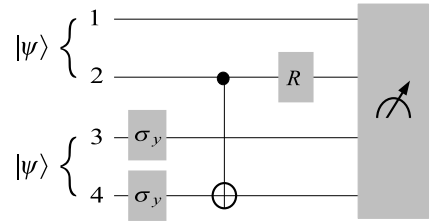}
\caption{Quantum circuit describing a direct measurement of the
concurrence of a two-qubit pure state, where two copies are
available. It involves a controlled-not gate, as well as
$\sigma_{y}$ unitaries and other simple $R$ qubit rotations,
followed by the joint measurement of the four
qubits.}\label{protocol}
\end{figure}

Let us assume that we want to measure the concurrence of the general
two-qubit pure state
\begin{equation}
\label{psi} \left\vert \psi \right\rangle =c_{0}\left\vert
gg\right\rangle +c_{1}\left\vert ge\right\rangle +c_{2}\left\vert
eg\right\rangle +c_{3}\left\vert ee\right\rangle ,
\end{equation}
and we are provided with two decoupled copies of it $\left\vert \psi
\right\rangle \otimes \left\vert \psi \right\rangle$. It can be
shown from Eq.~(\ref{pureconcurrence}) that the concurrence of state
$| \psi \rangle$ in terms of coefficients $c_i$ is given by
$C(\left\vert \psi \right\rangle )=2\left\vert
c_{1}c_{2}-c_{0}c_{3}\right\vert$. Following the suggested quantum
circuit of Fig.~\ref{protocol}, we apply local operations on the
second copy such that the global state is described by $| \Phi
\rangle = \left\vert \psi \right\rangle \otimes \left( \sigma
_{y}\otimes \sigma _{y}\left\vert \psi \right\rangle \right)$. This
state can be written as a superposition of states having a number
excitations $k$ ranging from $0$ to $4$,
\begin{equation}
\begin{tabular}{l}
$\left\vert \Phi \right\rangle =-c_{0}c_{3}\left\vert gggg\right\rangle $ \\
$\ \ \ \ \ \ \ \ \ +c_{2}c_{0}\left\vert ggge\right\rangle
+c_{0}c_{1}\left\vert ggeg\right\rangle $ \\
$\ \ \ \ \ \ \ \ \ -c_{1}c_{3}\left\vert gegg\right\rangle
-c_{2}c_{3}\left\vert eggg\right\rangle $ \\
$\ \ \ \ \ \ \ \ \ -c_{0}^{2}\left\vert ggee\right\rangle
-c_{3}^{2}\left\vert eegg\right\rangle +c_{1}c_{2}\left\vert
gege\right\rangle $ \\
$\ \ \ \ \ \ \ \ \ +c_{1}^{2}\left\vert geeg\right\rangle
+c_{2}^{2}\left\vert egge\right\rangle +c_{2}c_{1}\left\vert
egeg\right\rangle $ \\
$\ \ \ \ \ \ \ \ \ -c_{1}c_{0}\left\vert geee\right\rangle
-c_{2}c_{0}\left\vert egee\right\rangle $ \\
$\ \ \ \ \ \ \ \ \ +c_{3}c_{2}\left\vert eege\right\rangle
+c_{3}c_{1}\left\vert eeeg\right\rangle $ \\
$\ \ \ \ \ \ \ \ \ -c_{3}c_{0}\left\vert eeee\right\rangle$ .
\end{tabular}
\end{equation}
Now, we apply a CNOT operation between the second qubit acting as
the control and the fourth qubit acting as the target, followed by a
rotation on the second qubit. The CNOT gate in this protocol is
defined such that if the control qubit is in state $\left\vert
g\right\rangle $ the target is not affected, conversely, if the
control is in the state $\left\vert e\right\rangle $ the target is
flipped. The subsequent rotation $R^-_2$ acting on qubit $2$ can be
taken from $R^{\pm}_j : \left\vert g \right\rangle _{j} \rightarrow
( \left\vert g \right\rangle _{j} \pm \left\vert e \right\rangle_{j}
) / \sqrt{2}$ and $\left\vert e \right\rangle_{j}\rightarrow (
\left\vert e \right\rangle_{j} \mp \left\vert g \right\rangle_{j} )
/ \sqrt{2}$. After the CNOT and $R^-_2$ operations, the state of the
overall system becomes

\begin{equation}
\begin{array}{lll}
\left\vert \Phi _{1}\right\rangle =\frac{1}{\sqrt{2}} & \{ & A_{-}\left\vert
gggg\right\rangle +A_{+}\left\vert gegg\right\rangle \\
& + & B_{-}\left\vert ggge\right\rangle -B_{+}\left\vert
gege\right\rangle
\\
& + & 2c_{2}c_{3}\left\vert eegg\right\rangle
-2c_{0}c_{1}\left\vert
geeg\right\rangle \\
& + & C_{10}^{-}\left\vert ggee\right\rangle +C_{10}^{+}\left\vert
geee\right\rangle \\
& + & C_{23}^{-}\left\vert egge\right\rangle -C_{23}^{+}\left\vert
eege\right\rangle \\
& + & A_{-}\left\vert egeg\right\rangle -A_{+}\left\vert
eeeg\right\rangle
\\
& + & B_{+}\left\vert eeee\right\rangle -B_{-}\left\vert
egee\right\rangle
\},%
\end{array}
\label{fi1}
\end{equation}
where $A_{\pm }=c_{1}c_{2}\pm c_{0}c_{3}$, $B_{\pm }=c_{0}c_{2}\pm
c_{1}c_{3} $, and $C_{ij}^{\pm }=c_{i}^{2}\pm c_{j}^{2}$. We observe
that in Eq.~(\ref{fi1}) the concurrence information of state
$\left\vert \psi \right\rangle$ is present in the coefficient
$A_{-}$ through
\begin{equation}
\label{theequation} C(\left\vert \psi \right\rangle
)=2\sqrt{2P_{gggg}},
\end{equation}
where $P_{gggg} = |A_{-}|^2 / 2$. Clearly, a similar argumentation
leads also to $C(\left\vert \psi \right\rangle )=2\sqrt{2P_{egeg}}$.
We will clarify our choice below when discussing applications to
specific experimental setups.

We consider now the proposed protocol for the case of atoms flying
through 3D microwave cavities, an important physical setup where
fundamental tests of quantum mechanics have been
realized~\cite{raimond01}.

\begin{figure}[t]
\includegraphics[width=80mm]{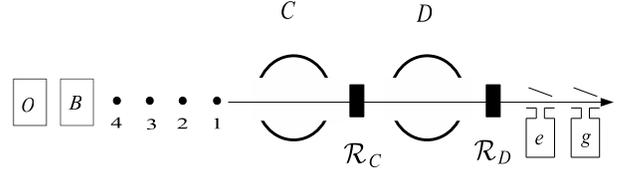}
\caption{Protocol for measuring concurrence in microwave 3D cavity
QED using two cavities and two Ramsey regions. } \label{cavity}
\end{figure}

The proposed protocol will make use of two cavities, two Ramsey
regions, and Rydberg atoms crossing them at given velocities, see
Fig.~\ref{cavity}. It relies on present efforts to develop
two-cavity setups~\cite{privatecomm1}, but see also other
multi-cavity projects~\cite{paperfriends,varcoedream}. The first
cavity $C$ is used to create two copies of the same entangled
two-atom state in a consecutive manner. We make use of an entangling
technique that has already been experimentally demonstrated in
Ref.~\cite{haroche2001}, following the proposal of
Ref.~\cite{zheng}. Along these lines we are entitled to say that a
general entangled state of the form $\alpha |g e \rangle + \beta | e
g \rangle$ could be produced in the lab. In Ref. \cite{haroche2001},
two Rydberg atoms, with a relative delay $\tau$, are sent from $B$
with velocities $v$ and $w$ ($w >v$) such that they cross inside the
cavity, determining in this way the desired effective Rabi angle. We
propose here to create the two required copies one after the other,
where the atoms of each pair will have the same velocities $v$ and
$w$, and a suitable delay time $\tau'$ between atoms $2$ and $3$. We
will see below that while requiring the atom pairs $\{ 1 , 2 \}$ and
$\{ 3 , 4 \}$ to cross inside cavity $C$, for generating the same
entangled state $| \Psi \rangle$, atoms $2$ and $4$ will not need to
cross in $D$ to produce the CNOT gate.

Before cavity $C$, see Fig.~\ref{cavity}, the four atoms follow the
natural order $\{ 4,3,2,1 \}$, from left to right. Immediately after
cavity $C$, the four atoms encoding the initial state $| \Psi
\rangle \otimes | \Psi \rangle$, follow the ordering $\{ 3,4,1,2 \}$
due to the timing and velocities mentioned above. To begin with the
protocol described in Fig.~\ref{protocol}, we allow now, atoms $4$
and $3$ to cross the Ramsey region where local unitaries $\sigma_y$
are applied. We consider that Ramsey regions were not active when
atoms $2$ and $1$ passed through at an earlier time. We recall that
Ramsey zones, implementing different local rotations are
well-controlled and accurate devices, representing an important
building block of present technology in 3D microwave
cavities~\cite{raimond01}. Note that short before entrying cavity
$D$, it would be preferable to have the following ordering $\{
3,1,4,2 \}$. This exchange of positions between atoms $1$ and $4$
could be easily achieved by proper tuning of parameters $v$, $w$,
$\tau$, $\tau'$, and the distance between cavities.

The second step of the protocol is the implementation of a CNOT(2,4)
gate between control atomic qubit $2$ and target atomic qubit $4$.
As explained before, atom $2$ arrives first to cavity $D$ followed
by atom $4$. It can be easily proved that this gate is equivalent to
the successive operations $R^+_4 \times {\rm CPHASE(2,4)} \times
R^-_4$. The CPHASE(2,4) gate acts as follows: $| e \rangle_2 | e
\rangle_4 \rightarrow - | e \rangle_2 | e \rangle_4 $, while the
other basis states, $\{ | g \rangle_2 | g \rangle_4 , | g \rangle_2
| e \rangle_4 , | e \rangle_2 | g \rangle_4 \}$, remain unchanged.
To achieve this goal we map first the qubit of atom $2$ onto the
photonic state of cavity $D$. Then, atomic qubit $4$ is transformed
due to $R^-_4$ and enters into cavity $D$ to perform a CPHASE with
the photonic qubit, that is, $| e \rangle | 1 \rangle \rightarrow -
| e \rangle | 1 \rangle$, leaving other states unchanged. We suggest
the use of the CPHASE gate implemented experimentally in
Ref.~\cite{haroche1999}. Along these lines, we propose the use of an
upper auxiliary level $| i \rangle$ allowing a $2\pi$-pulse rotation
in the subspace $\{ | e \rangle | 1 \rangle , | i \rangle | 0
\rangle \}$~\cite{comment}. Finally, atomic qubit 4 is rotated
through the action of $R^+_4$, while the photonic qubit is mapped
back onto an additional atom $5$ in its ground state.

As is evident from above, atom $2$ is lost in this process but its
logical information is carried now by atom 5. A last step consists
on measuring the level statistics of all qubits after a final
rotation $R^-_5$ is implemented on atomic qubit $5$, following the
protocol of Fig.~\ref{protocol}. As shown in
Eq.~(\ref{theequation}), the probability of finding all relevant
atoms $\{ 5,3,1,4 \}$ in the ground state will provide us with a
valuable information: the concurrence of the entangled pure state $|
\Psi \rangle$. Clearly, following Eq.~(\ref{fi1}), we can obtain
similar information if we use the probability $P_{egeg}$.

There are additional technical points in order. First, it would be
desirable that atom $5$ is sent with the proper timing and velocity
so that it can retrieve the photonic qubit of cavity $D$ before
decoherence destroys the encoded information of atom $2$. Second,
the proper tuning of the relative frequency of cavities $C$ and $D$
can only be decided when all experimental parameters, including
inter-cavity distance and atomic transitions, are decided. Third, we
rely our proposal on the possibility of switching on and off at any
desired time the Ramsey regions, as well as in controlled DC-induced
shifts in the atomic transition frequencies~\cite{dcpulses}.

\begin{figure}[t]
\includegraphics[width=80mm]{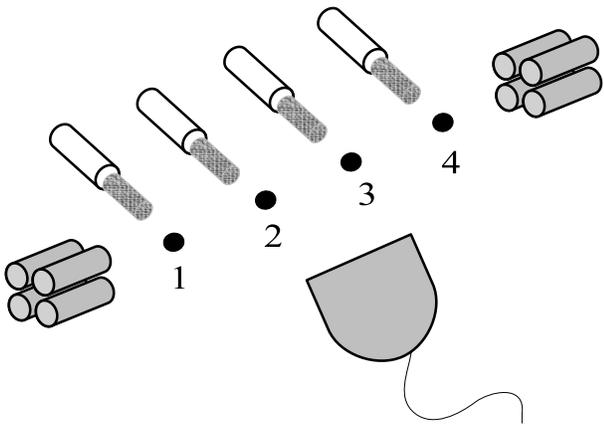}
\caption{Four ions in a linear Paul trap that can be individually
addressed, measured, and coupled to a collective motional degree of
freedom to implement the protocol of Fig.~\ref{protocol}.}
\label{ions}
\end{figure}

Alternatively, the protocol of Fig.~\ref{protocol} could be
implemented straightforwardly in four trapped ions, see
Fig.~\ref{ions}, as will be discussed below. For achieving that goal
we require to implement $\sigma_y$ unitaries, local rotations
$R^{\pm}$, and a CNOT gate, all of which have already been
implemented in the lab with high precision in several trapped ions.
That is, we rely on the possibility of implementing individual
addressing on each of the four ions, for the sake of individual
control and readout. Typically, the measurement of each ion is done
by means of an electron-shelving technique, where an internal level,
say $| e \rangle$, is coupled to an auxiliary level $| c \rangle$
that decays cyclicly back to $| e \rangle$. The abundance of
fluorescence photons implies the projection of the qubit on state $|
e \rangle$, and the absence of fluorescence photons warrants the
projection of the qubit on the other state $| g \rangle$. However,
we remark that, for measuring the concurrence according to the
proposed protocol, it is not necessary to realize an individual
readout of the ionic qubits. We propose here the use of a technique
that may be called {\it global
electron-shelving}~\cite{twoshelving}, where the required
measurement of $P_{gggg}$ of Eq.~(\ref{theequation}) is reduced to a
single-bit yes/no question. Given that all ions are identical, we
can apply the same electron-shelving laser pulse globally and
simultaneously, but each ion will perform its individual associated
cyclic transition. Only the absence of fluorescence photons warrants
the projection of the four-qubit state onto state $| g \rangle | g
\rangle | g \rangle | g \rangle$, while the presence of fluorescence
photons implies a projection on any other four-qubit state. It may
even happen that, while implementing the yes/no global photon
statistics, the multi-ion case produces a higher fidelity in the
desired probability measurement when compared to the individual
readout case. In this manner we would accomplish the measurement of
the concurrence through a simplified and global readout scheme for
$P_{gggg}$ of Eq.~(\ref{theequation}).

In conclusion, we have presented a realistic protocol for measuring
directly the concurrence of a two-qubit pure state in matter qubits,
as long as two copies and a few simple operations are available. We
have shown that it can be applied in a two-cavity setup in microwave
3D cavity QED and straightforwardly in trapped ion systems. We
believe that this proposal could be implemented with present
technology and will boost research in the hard task of quantifying
entanglement of small dimensional systems.

The authors thanks T. Sch\"atz for useful discussions. GR
acknowledges financial support from CONICYT Ph.D. Programm
Fellowships, CEL and FL from MECESUP USA0108, JCR from Fondecyt
1030189 and Milenio ICM P02-049, and ES from EU RESQ, EuroSQIP, and
DFG SFB 631 projects. CEL thanks also DIGEGRA USACH and Jan von
Delft for hospitality at LMU.

\end{document}